# Algorithm Based on One Monocular Video Delivers Highly Valid and Reliable Gait Parameters

Dr. Arash Azhand¹*, Dr. Sophie Rabe¹, Dr. Swantje Müller¹, Igor Sattler², Dr. Anika Steinert²

*corresponding author: arashazhand@gmail.com

¹ Lindera GmbH, Kottbusser Damm 79, 10967 Berlin, Germany

² Geriatrics Research Group, Charité–Universitätsmedizin Berlin, corporate member of Freie Universität Berlin, Humboldt-Universität zu Berlin, and Berlin Institute of Health, Charitéplatz 1, 10117 Berlin, Germany

## Abstract

Despite its paramount importance for manifold use cases (e.g., in the health care industry, sports, rehabilitation and fitness assessment), sufficiently valid and reliable gait parameter measurement is still limited to high-tech gait laboratories mostly. Here, we demonstrate the excellent validity and test-retest repeatability of a novel gait assessment system which is built upon modern convolutional neural networks to extract three-dimensional skeleton joints from monocular frontal-view videos of walking humans. The validity study is based on a comparison to the GAITRite pressure-sensitive walkway system. All measured gait parameters (gait speed, cadence, step length and step time) showed excellent concurrent validity for multiple walk trials at normal and fast gait speeds. The test-retest-repeatability is on the same level as the GAITRite system. In conclusion, we are convinced that our results can pave the way for cost, space and operationally effective gait analysis in broad mainstream applications. Most sensor-based systems are costly, must be operated by extensively trained personnel (e.g., motion capture systems) or – even if not quite as costly – still possess considerable complexity (e.g., wearable sensors). In contrast, a video sufficient for the assessment method presented here can be obtained by anyone, without much training, via a smartphone camera.

## Introduction

The quest to understand human gait patterns and their impact on human health and well-being is a very old scientific endeavor, spanning multiple disciplines, and closely associated with the development of the scientific method **[1-4]**. In this vein, the systematic study of human gait has been continuously refined over the centuries, up to our modern times **[1-3, 5-13]**.

The advent of computers made it possible to process large amounts of data much more efficiently, allowing for the developing of gait assessment laboratory systems. However, these systems mostly consist of 4 to 10 video cameras in a lab setting. Hence, they are very difficult to setup, calibrate and operate [1, 14, 15]. Moreover, the inherent complexity of combining multiple cameras and body joint markers means that measurement accuracy and precision are considerably dependent on the positioning and calibration of the system [14, 15].

One other popular gold-standard tool for clinical gait assessment is the GAITRite pressure-sensitive carpet system (hereafter referred to as GS), which has been validated against the Vicon system [16, 17]. Though this system exhibited high validity and repeatability, it is still not accessible to the broad mainstream. A standard GS, 5.2m in length, comes at a price point of around 40.000 Euro and needs trained personnel to be operated correctly.

Given the need and demand of precise and reliable human motion analysis in diverse areas, it is paramount to assess if modern computer vision algorithms (classification, detection, and segmentation of objects in images) might have the potential for powering state-of-the-art human mobility and gait assessment systems of the future at much lower cost in financial and human resources compared to traditional motion capture systems and gait labs.

Within the last decade a novel class of data-based learning algorithms on the basis of neural networks revolutionized the area of computer vision. Since the seminal paper by Yann LeCun et al. In 1998 [18], Deep Learning (DL) methods in general and such image-based Convolutional Neural Networks (CNNs) in particular helped solve more and more previously assumed hard to unsolvable problems in computer vision [19]. The year 2012 marked the date after which DL based methods took the lead over traditional computer vision methods. At that time Alex Krizhevsky, Ilya Sutskever, and Geoffrey Hinton won with their CNN model [20] the 2012 ImageNet Large-Scale Visual Recognition Challenge (ILSVRC). Their results in the classification task were at 15.3% error rate compared to the 26.2 % of the second-best team [21]. This was the starting point for the development of better and more accurate CNN models within just the last decade [22-25].

In parallel to the development of more and more accurate CNN models for computer vision tasks in general, the special problem of human pose estimation made huge leaps. In essence, CNN models serve the backbone of many of the most successful human skeleton pose estimation algorithms of the last years up until now. Some examples for human pose estimation in 2D include the Convolutional Pose Machine (CPM) [26] that is utilized within the OpenPose 2D detector [27], the DeepCut model [28], RMPE (AlphaPose) model [29], and the most recent HRNet model [30]. In addition, many DL based approaches in estimating skeleton joints in 3D space have been pursued, some of which are estimating 2D and 3D poses end-to-end from the video frames [31-37]. Recently, another type of approach in 3D pose estimation, that is termed 3D uplifting and is taking existing 2D skeleton data instead of the video frames as input, has also shown wide success [38-40]. In this case, one of the 2D pose estimation algorithms can serve as the detector which extracts the 2D input for the 3D uplifting model from the input video.

The era of DL based computer vision algorithms for human pose estimation promises many applications in diverse areas such as the health care [41-45], sports [46-50], and entertainment [51-57]. Particularly, in case of health care, the outcome parameters that are estimated or computed based upon such extracted human skeleton joints, must hold to a high accuracy and reliability standard. Medical decisions might be depending on the level of delivered accuracy. Mostly, very complex and costly motion capture systems or other Gold Standard gait assessment systems are utilized in the medical area.

Some recent approaches were undertaken to show the performance of more mobile systems. Steinert et al compared two single camera systems with GS [58]. But one of the studied systems in that study is based on the Microsoft Kinect version 2 and as such not as mobile as a Smartphone based system, though the validity and repeatability of the delivered gait parameters in comparison to GS were shown to be at least on medium to high level. The second studied system was a Smartphone system, but the underlying DL algorithm was only able to deliver gait parameters with low to medium level validity and repeatability measures. Kidziński and colleagues trained a ML algorithm for estimating gait parameters from trajectories of 2D body poses extracted from videos using OpenPose [59]. Apart from the fact that that system is utilizing only a 2D pose estimation system, it is just trained on solely a set of 1792 videos from 1026 unique patients with cerebral palsy. Thus, to our opinion such a system will be very prone to over-fitting to very particular group, since learned on a very specific class of patients. Moreover, gait parameter estimation on the basis of just 2D skeleton pose trajectories are very error-prone, since there is a considerable amount of information loss when projecting from real 3D poses to 2D poses. Even more recently, Stenum and colleagues also used 2D pose estimation data (OpenPose) to extract gait parameters [60]. As expected, due to the limitations of utilizing just 2D human estimation data, the validity results were shown to be very mixed (low to good level).

Contrary to previous studies, in this publication we study the validity and repeatability of a novel system on the basis of monocular videos. The novelty of that assessment system is that it directly and in a mathematically deterministic way calculates the gait parameters from fully three-dimensional skeleton poses. The first two modules estimate the 2D and the 3D skeleton poses, respectively, via modern CNNs for each frame of the input video. The third module takes these 2D and 3D skeleton pose data for all frames and optimizes the 3D skeletons according to the given height of the person in video. In addition, the optimizer also estimates the 3D coordinates of the camera in the skeleton centered space. The final module of the pipeline is a patented [61] mathematical algorithm that functionally calculates the gait parameters from the anatomically optimized 4D skeleton data (3 spatial coordinates plus the time) and the camera position in space. The technical details of this algorithm pipeline, that hereinafter is referred to as the smartphone camera application (SCA), will be provided in the Methods section.

In the next section, we present and discuss the results of the validation study. We then provide a summary of the main findings. These findings will be explained thoroughly within the Discussion section. Finally, details concerning the assessment systems (GS and SCA) and the data generation methods are provided in the Methods section.

# Results

We performed gait assessment for the study with the GS and SCA. The results presented and discussed here were achieved on the basis of 514 walking videos acquired from 44 healthy adults with an average age of 73.9 years (± 6.0 years, range: 65 to 91 years). The subjects were asked to perform three walks at their preferred walking speed and subsequently three walks at a fast speed. All walks were performed on the GS. Simultaneously, two videos were captured for each walk sequence by each subject, one using a hand-held smartphone (SCA Hand) and one using a smartphone fixed on a stand (SCA Stand).

Table 1 summarizes the outcomes of the statistical analysis. The mean and standard deviations for both SCA Hand and SCA Stand do not differ significantly from the reference system GS. We calculated the mean differences between SCA (both Hand and Stand) and GS as a measure of bias between the systems and transformed these mean differences (Diff) in percentage of the GS means.

This yields 0.2% percentage bias for step length to 0.7% for gait speed in case of SCA Hand. In contrast, the percentage mean differences for SCA Stand are between 0.6% for step time up to 2.8% for gait speed. Fig. 1 shows the distribution plots for these mean differences with 95% CI for all measured gait parameters.

This affirms that no gait parameter measured with the SCA differs from the corresponding GS measurements by more than 3.25% of the corresponding gait parameter mean. We calculated these plots and the values presented in Table 1 for Diff [95% CI] using bootstrapping techniques [62], since the underlying distributions of the differences are unknown as well as skewed.

The Inter-class correlation coefficients [63, 64], ICC(2, k), range between 0.958 and 0.987 for all gait parameters. The ICC is a measure of the agreement between two measurement methods and is defined as the ratio of the variance between –1 and +1. An ICC of 0 indicates random evaluation behavior, a value of 1 an ideal reliable feature evaluation by the method. According to [64], ICC values above 0.9 can be considered as excellent agreement.

The repeatability of the SCA conditions (SCA Hand and Stand) is measured by the ICC (3, 1). These values range between 0.915 and 0.950. Previous studies on the validity of the GS found it to possess ICC (3,1) measures between 0.84 and 0.97 for the same two walking conditions [16]. The ICC (3, 1) values of parameters as measured with GS for our dataset are between 0.91 and 0.98 (not shown here). Thus, the repeatability of the SCA is on the same level as GS.

Fig. 2 displays the Bland-Altman plots for gait speed, cadence, average step length, and average step time. Each plot is based on both types of walking speed trials. There is one plot each for SCA Hand and SCA Stand. Each single blue dot in the plots indicate an average value along one walk trial (one video). The Bland-Altman plots also contain the corresponding mean absolute differences, as a measure of the actual measurement error between the systems, together with the lower and upper 95% CI. Again, the mean absolute differences are between 0.5% and 2% of corresponding gait parameter mean values. The lower and upper 95% CI as a percentage of the corresponding mean values are the narrowest for step time and cadence, with an average of -5% on the lower interval up to +7% on the upper interval. The widest 95% CI intervals are observed for gait speed and step length, from –11% to +13%.

**Table 1:** Statistical comparisons between GS and SCA (Hand and Stand) averaged over preferred and fast gait speed trials. Means and standard deviations (SD) are reported together with inter-class-correlation coefficients, ICC (2, k), for multiple gait parameters, inter-trial repeatability as measured by ICC (3, 1), and measurement differences between the two systems (GS and SCA) in terms of mean values with lower and upper bounds of confidence intervals (Diff [95% CI]) transformed to percentage of means.

|  | GS | SCA Hand |  |  |  | SCA Stand |  |  |  |
| --- | --- | --- | --- | --- | --- | --- | --- | --- | --- |
|  | Mean (SD) | Mean (SD) | ICC (2, k) | ICC (3,1) | Diff [95% CI] in % of mean | Mean (SD) | ICC (2, k) | ICC (3,1) | Diff [95% CI] in % of mean |
| Gait speed [m/s] | 1.42 (0.32) | 1.41 (0.31) | 0.982 | 0.950 | 0.28 [-0.40; 1.01] | 1.38 (0.32) | 0.981 | 0.935 | 2.5 [1.83; 3.25] |
| Cadence [steps/min] | 121.77 (14.56) | 121.15 (13.44) | 0.983 | 0.930 | 0.52 [0.15; 0.88] | 120.76 (13.14) | 0.980 | 0.915 | 0.84 [0.45; 1.21] |
| Step length [cm] | 69.22 (10.17) | 69.37 (10.39) | 0.961 | 0.950 | 0.23 [-0.92; 0.48] | 68.01 (10.63) | 0.958 | 0.930 | 1.77 [1.04; 2.50] |
| Step time [s] | 0.500 (0.061) | 0.502 (0.057) | 0.987 | 0.925 | -0.36 [-0.68; -0.04] | 0.503 (0.056) | 0.984 | 0.915 | -0.62 [-0.95; -0.26] |

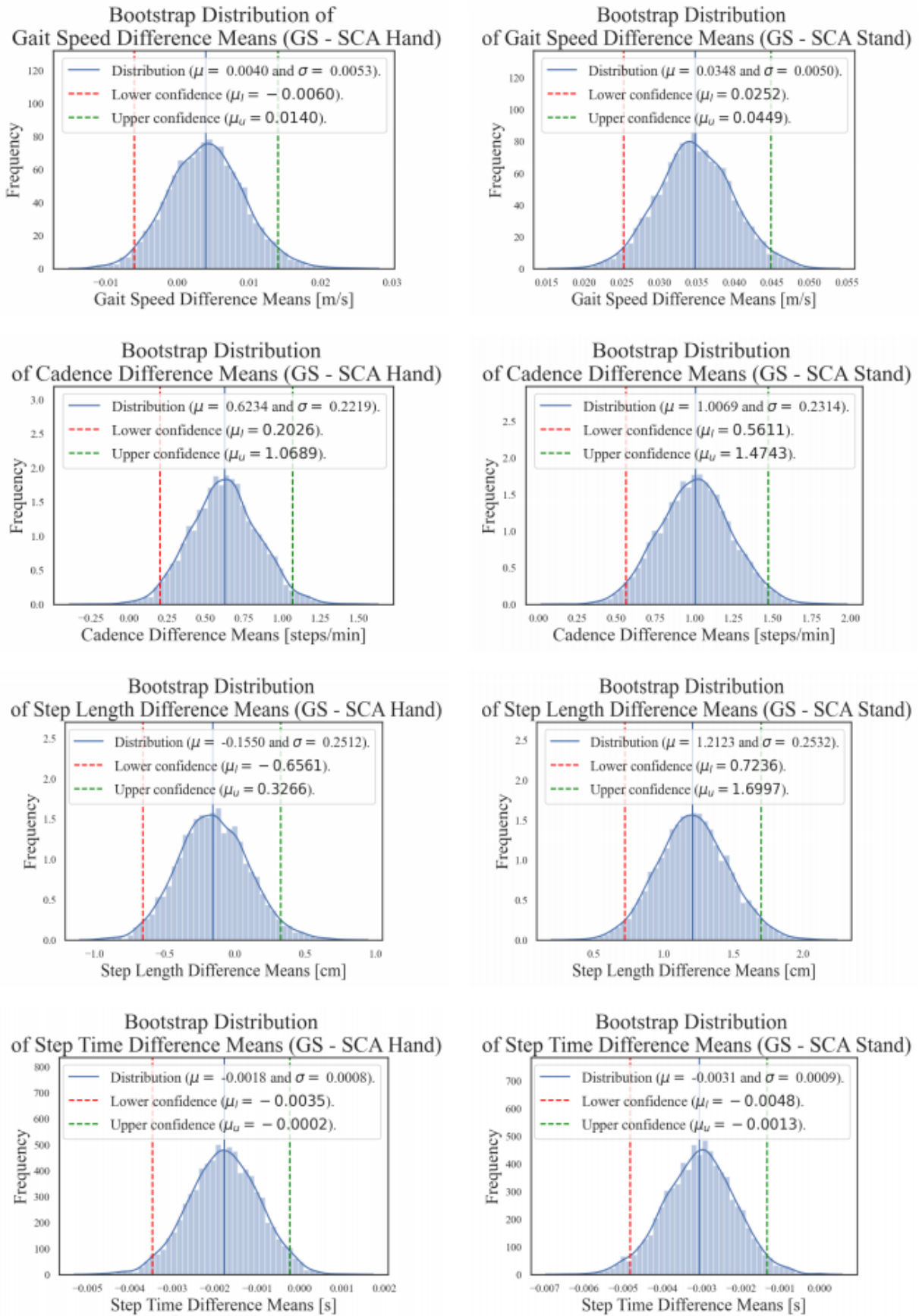

**Fig. 1:** Mean differences with 95% CI values calculated via bootstrapping for each measured gait parameter in the SCA conditions (Hand and Stand) as compared to the GS.

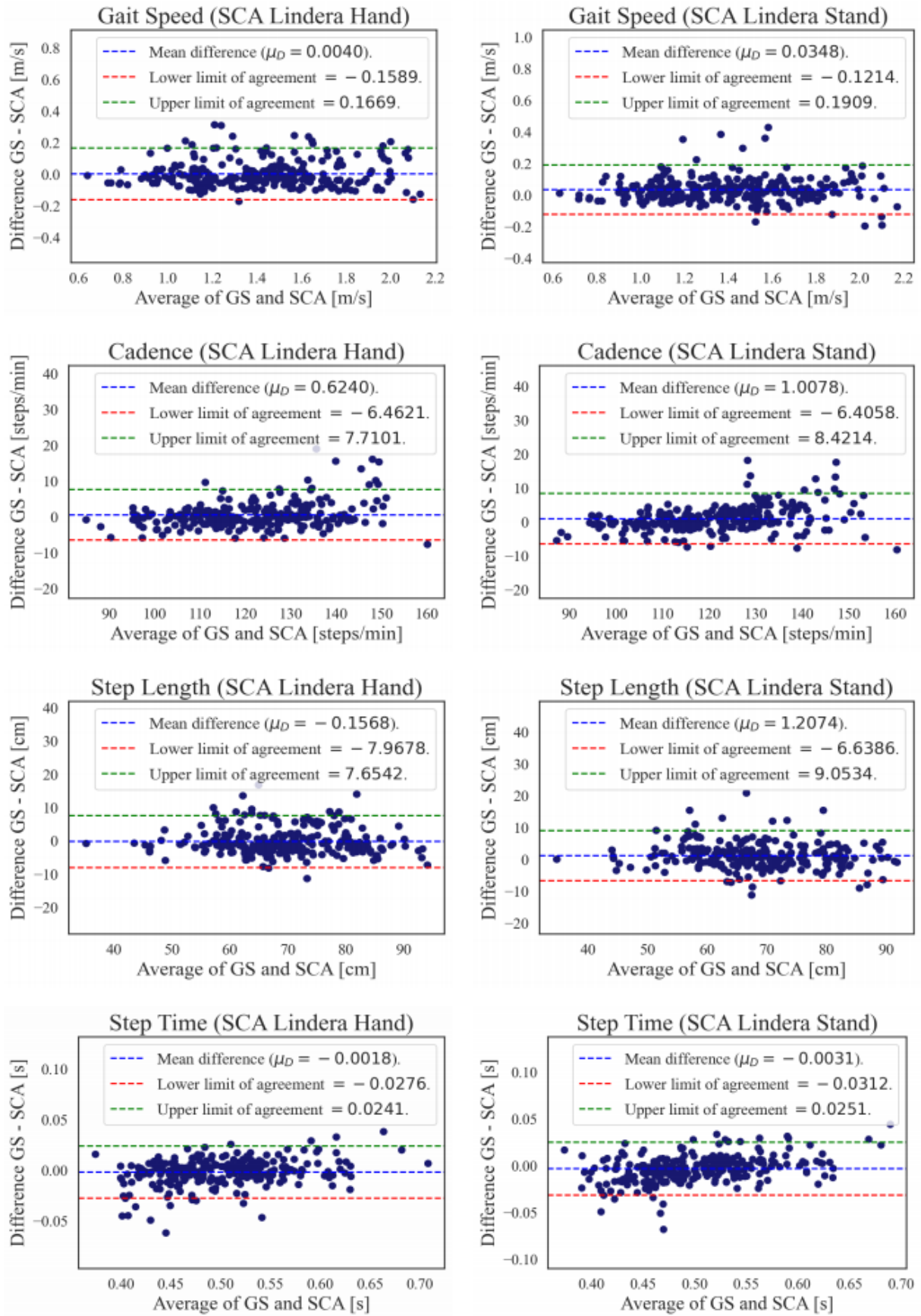

**Fig. 2:** Bland-Altman plots for each measured gait parameter and SCA conditions (Hand and Stand)

# Discussion

The purpose of this study was to demonstrate the validity of a purely algorithmic gait assessment application based on walking videos taken with monocular smartphone cameras (smartphone camera application, SCA) in relation to the Gold Standard gait assessment system GAITRite (GS). Purely algorithmic means here that extraction of the gait parameters does not depend on a specific device (e.g., smartphone). The videos used in the study were acquired through the SCA using a smartphone camera, but the underlying algorithm pipeline for extracting the gait parameters from the walking videos is independent of the application and the underlying device. Hence, there is no added complexity due to device-specific calibration, sensor placement or the calibration and synchronization of motion-capture systems.

All statistical measures showed excellent validity of SCA. In particular the ICC (2, k) measures are between 0.958 and 0.987 for all parameters, at both walking conditions, and for both smartphone devices (Hand and Stand). The Bland-Altman analysis conducted in this study provided insight into the overall measurement error for SCA. We found the error compared to the GS to be between 5% and 13% of the corresponding gait parameter mean value. To better understand the meaning and significance of these error measures, they must be set in relation to other assessment system outcomes as well as the threshold below which specific percentage errors are considered clinically acceptable. According to **[65]**, clinical acceptability corresponds to a percentage error below 30%. Moreover, two other very recent approaches to gait assessment exhibited percentage errors significantly higher than the approach presented here. Mun et al use a foot-feature measurement system and then estimate the gait parameters directly via a neural network **[66]**. Their percentage errors are between 15% and 30%. Another recently published approach compares a system called Smart Walker with the GS **[67]**, yielding mixed results with respect to percentage error (8.7 - 23.0% for stride, swing and stance time and 31.3% to 42.3% for stride length and gait speed). In contrast, even the highest percentage errors for the SCA (11 to 13.5% for gait speed and step length) are similar to or even below the lowest values of the other two recent approaches. The SCA's test-retest repeatability, measured via ICC (3, 1) are on the same level as the GS.

In conclusion, to the best of our knowledge, the approach studied here exhibits the best validity and repeatability of a monocular video-based skeleton tracking system in comparison to a gold-standard gait assessment system and therefore be considered state-of-the-art. It should be noted that these outcomes were achieved across both walking conditions (fast and preferred) and irrespective of measurement condition (SCA Hand and Stand).

On the other hand, there are still some shortcomings to discuss. For example, there is an observable higher error rate in step length (> 10 cm in some cases) visible explicitly in the Bland-Altman plot (Fig. 2). Moreover, the measurement by SCA Stand seem to be slightly worse in some cases in comparison to SCA Hand, although the former is mounted on a fixed stand. To our suggestion one reason for these observations might be the fact that during the data acquiring sessions (see in the Methods section, particularly Fig. 5), the SCA Stand was placed slightly inclined relative to the walk way. Hence, a reason for slightly worse results in step length by SCA Stand could be due to potential errors arising at the edges of the field of view, because of parallax and changes in perspective **[60]**. Additional sources of error in general are rooted at the center of the CNN based skeleton detector algorithms utilized. Although these models improve in accuracy and stability nearly monthly to quarterly, there are still some issues in reliably detecting skeletons in specific environmental situations like clothing, background, lighting, etc. Moreover, lower body parts are still not estimated as reliably compared to other parts because of lack of data for these parts.

Another specific limitation of the employed algorithm pipeline herein is its sole current applicability on frontal view videos, though it is not specifically needed that persons should only walk to the camera (the walking back is also applicable). Additionally, the person making the video from the walking person should not walk to or away from the walking person simultaneously, since the gait parameter calculation relies on the stationary camera position. Otherwise, when the camera is moving, the validity and reliability of measured gait parameter values will deteriorate.

As an outlook, we are convinced that algorithmic computer vision based mobility assessment systems on single monocular videos are already on the verge to show their great potential for manifold application areas (medical, sports, entertainment and arts). Neural network based human pose estimation methods have been developed less than a decade ago and they are already able to provide basic backbones for highly valid and reliable gait assessment systems for a mere fraction of financial and other resources needed for multi-video and sensor-based gait laboratories. In addition, development in the field of computer vision and human pose estimation progresses at a breathtaking pace. Less than one and a half year ago the mean per joint error rate of state-of-the art 3D pose estimators were at a range of ~ 9 - 13 cm **[31]**. Today, current state of the art 3D uplifting models already show mean per joint error rates of ~2 – 3 cm **[38-40]**. With that progress in research and development together with gathering of more big data amounts (millions to billions of pose images with 2D and 3D ground truth labels), the perspective is on the verge to error rates into millimeter range. Furthermore, the huge amount of data introduces more variability across environmental situations (clothing, background, lighting, etc.) and also will improve the reliability of measurement for all body parts. The accuracy is just one potential progress of the models. Another huge potential concerns the run time on devices (e.g, mobile phones). There are current state-of-the-art 2D estimators running real time (20 – 24 frames per second) on mobile phones **[69, 70]**. Combined with a state-of-the-art 3D uplifting model, that already is running real time, a potential gait and mobility assessment system is imaginable, running on device with high accuracy and reliability on low-cost mobile devices everywhere on the planet, even without internet access.

# Methods

## Assessment Systems

Gait assessment in this study was conducted with two different assessment systems: GAITRite System (GS) and Smartphone camera application (SCA).

The GS is a carpet of 5.2 m length (active length of 4.27 m) and width of 90 cm (60 cm active) **[58]**. The carpet contains 16 128 embedded sensors in a grid. The sensors are placed at a distance of 1.27 cm and are activated by mechanical pressure. GS allows the measurement of different temporal (e.g., step time, walking speed, cadence) and spatial (e.g., step and stride length) parameters. The carpet is connected to a computer via an interface cable. Prior to the gait analysis, the participant's age, weight, height, and leg length (right and left) had to be entered manually.

The SCA performs gait analysis by recording a video of the subject with the monocular smartphone camera and applying a modular algorithm pipeline to extract the gait parameters (Fig. 3). The modules are described in the following.

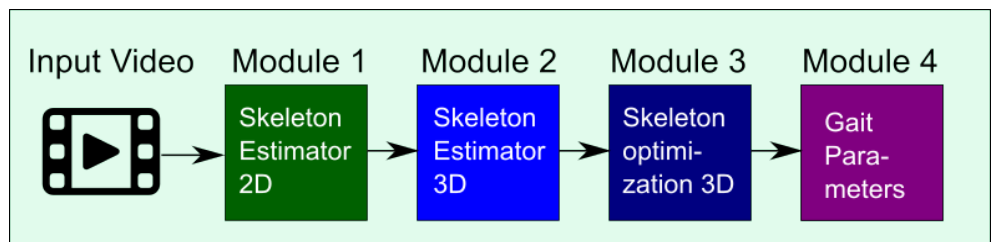

**Fig. 3:** Algorithm pipeline.

### Module 1 (2D Skeleton Tracker)

This module reads in the video sequence frame by frame and extracts a skeleton for each frame. These frame-wise skeletons are lists of 2D joint coordinates in the image pixel space. A sample is as follows:

{Neck: {x: 150, y: 50}, …, Left Hip: {x:170, y: 600}, … Left Ankle: {x: 175, y: 1000}, ...}.

The 2D tracker module has implemented our own variant of the Convolutional Pose Machine (CPM) 25 body joint model [26]. A sample image of a person with overlaid 2D skeleton is shown in Fig. 4. Moreover, our own variant of the HRNet model [30] is also implemented and can be switched in instead of the CPM 2D model.

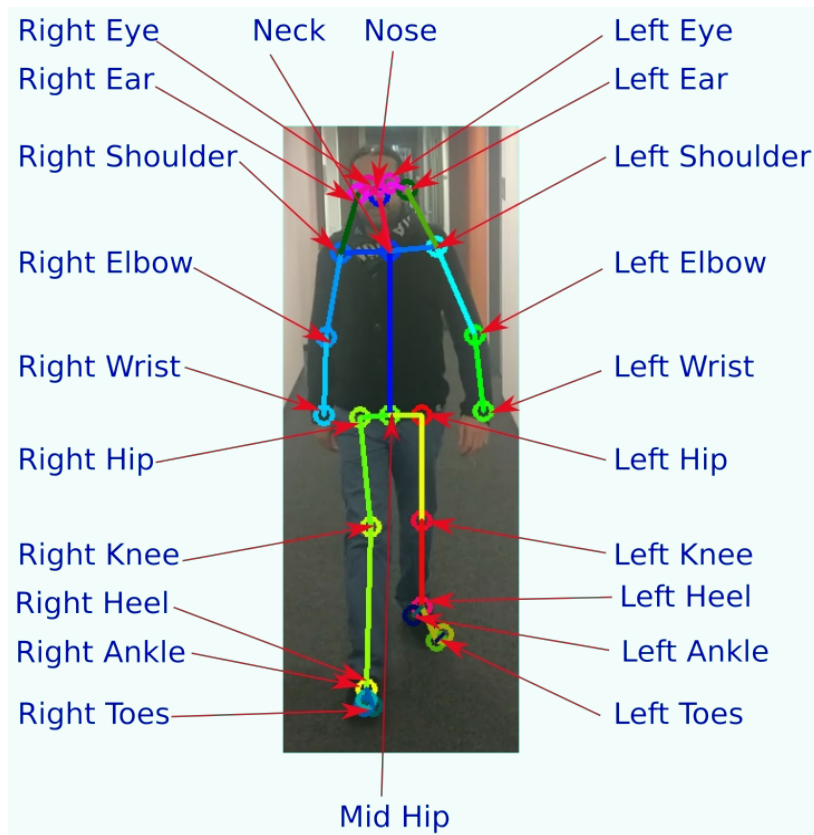

**Fig. 4:** Sample frame with estimated skeleton in 2D.

### Module 2 (3D Skeleton Tracker)

This module reads in either the video sequence frame by frame or it takes the 2D skeleton list frame by frame (output of a 2D detector) as input to then and extracts a list of 3D joint coordinates in meters for each frame. A sample is as follows:

{Neck: {x: 2.5, y: 1.9, z: 6.0}, …, Left Hip: {x: 2.1, y: 1.4, z: 6.0}, … Left Ankle: {x: 2.4, y: 0.12, z: 6.0}, ...}.

The 3D tracker module has implemented our own variant of the VNect model [31] and our own variant of the GASTNet 3D uplifting model [40]. Both the 2D and the 3D tracker modules are implemented in an abstract way, such that any state-of the art neural network model can be implemented as switchable tracker elements inside the pipeline modules.

### Module 3 (Skeleton Optimization 3D)

This module is our implementation of the kinematical skeleton fitter algorithm on top of the 3D detector [31]. It takes the previously estimated 2D and 3D skeletons together with the actual person height as inputs and then optimizes to the best fit 3D skeleton with correct anatomical bone lengths. The algorithm for optimization minimizes an energy objective function comprising four added terms. The first (inverse kinematics) term compares 3D joint coordinates that are inversely estimated from the 2D joints with the 3D joints estimated previously in module 2. The second (projection) term projects the estimated 3D skeleton back to 2D, then calculates the joint-wise errors between the projected and the estimated 2D skeleton.

In addition, the projection term determines global position related to the camera which gives us the per frame distance to camera. The third (smoothness) term applies temporal filtering in order to guarantee temporal consistency of motion between frames. Finally, the fourth and last term is the depth estimation term, which penalizes large variations in depth. The latter variations often arise because of large uncertainty in monocular reconstruction.

## Module 4 (Gait Parameter Calculation)

The gait parameters (gait speed, step length, step time, etc.) are directly calculated from the anatomically optimized 3D skeleton data. The latter consist of coordinates for the individual joints in space (x, y, z) over the entire video sequence. To determine the gait parameters from the sequence of these 3D joint coordinates, one can, in particular, determine the distance between the ankles for each frame of the video and then extract the extrema (maxima and minima), Theoretically, the individual moments in the video could be determined where a step is taken. However, the extraction of the actual extrema is not trivial, since the joint data is very noisy. A recently by us patented algorithm determines the correct extrema from a list of extrema candidates from this noisy skeleton data in the following manner [61]:

When the walking person is smoothly walking, the distance of one foot joint to a particular reference joint location changes periodically showing local maxima and local minima between two consecutive touch points of the respective feet with the ground. In general, a reference joint location is at best a further joint of the human body with respect to the foot location. If, for example, the reference joint is the joint coordinate position of the second foot, then a local maximum of the distance is reached each time when the moving feet has reached the step length, while the local minimum is reached at the moment when the second foot passes the location of the first foot [61].

In a real world scenario it may occur that multiple frames associated with a frame sequence where both feet are touching the ground show multiple local minima (maxima) of this distance. In this case, the system may determine a cluster of honest local maximum frames and/or honest minimum frames. In other words, honest local extrema frames include such frames which collectively reflect the distance between the feet for a particular step of the walking person. The average step length can then be determined by averaging the respective values for a plurality of honest local extrema (e.g., honest local maxima clusters) [61].

In summary, the process of gait parameter calculation is as follows. Firstly, the skeleton tracking delivers 4D skeleton data: per frame (time) a list of 3 spatial coordinate points (x, y, z) for 21 skeleton joints. Then, corresponding gait parameters are calculated step-by-step along the whole walk sequence, after we detected the specific left and right steps according to the patented algorithm [61]. The averages are then calculated from the step-by-step values, since the ground-truth from GS are only average values (validity comparison only possible with averages). In addition to the reported gait parameters in here, we are also calculating more gait parameters (e.g. foot heights) and any kinematical angle parameter (e.g., knee angles, hip frontal and side angles etc.). But, since GS just delivers solely ground level parameters (gait speed, step length, step time, cadence), we report validity study on these parameters here.

## Assessment Procedure

In the following we describe the assessment process in some detail [58]. The procedure for data generation was carried out in accordance with relevant guidelines and regulations. Firstly, an informed consent was obtained from all subjects. Both the ethics committee and the data protection office of the Charité-Universitätsmedizin Berlin approved all study procedures [58]. The assessment was conducted simultaneously with the GS and the SCA at the Geriatrics Research Group, Charité-Universitätsmedizin Berlin. The setup is illustrated in Fig. 5. The participants were recruited from former contacts of the Geriatrics Research Group. The inclusion criteria for participants were a minimum age of 65 years and participants' signed informed consent. Exclusion criteria were severe cognitive, sensory, or motoric disorders, and a legal representative. In an initial telephone screening, participants were informed about the study procedure, the inclusion and exclusion criteria were checked, and an appointment for the gait analysis was arranged (minimum of 24 h after screening).

The gait analysis was conducted in the laboratory of the Geriatrics Research Group in Berlin. Firstly, participants were once again informed about the study procedure and data protection. After giving their informed consent, participants had to complete a questionnaire with regard to socio-demographic data (age, sex, weight, height, leg length; self-deployed questionnaire).

Subsequently, each participant was asked to complete six walks with two speed conditions: preferred gait speed (PGS) and fast gait speed (FGS; instruction: a velocity that is faster than you would walk normally). For the walks, participants were instructed to walk over the carpet (GS) beyond the end. For each walk, GS was first started at the computer and then SCA-Hand and SCA-Stand were initiated, before the participant's walk, so that all three systems (GS; SCA-Hand, SCA-Stand) acquired the participant's gait simultaneously (Fig. 5). GS measurement was initiated by the first pressure contact to the carpet. After each walk, study personnel verified that the systems were measuring correctly. This procedure yielded 12 walking videos for each subject, 528 overall. But during the procedure two walks were not measured appropriately by the GS. Hence, these were excluded from analysis herein. Moreover, the videos from one subject were not uploaded correctly via the SCA. Hence, the corresponding 12 videos could not be processed. In summary, 514 videos by 43 subjects were included in this study.

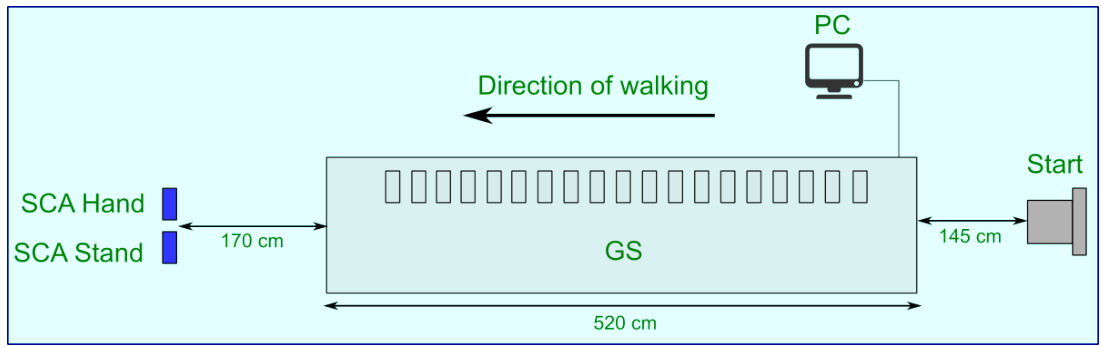

**Fig. 5:** Assessment setup.

## Statistical Analysis

To determine the level of agreement between the GS and SCA, inter-class correlation coefficients (ICC (2, k), two-way random effects, absolute agreement, single rater/measurement) were calculated **[63, 64]**. In addition, we calculated measurement differences between the two systems in terms of mean values and lower and upper bounds of confidence intervals (Diff [95% CI]) **[58]**. Moreover, Bland-Altman analysis **[68]** was performed to provide insights on the overall variability of the SCA in comparison to the GS. The mean differences with lower and upper limits of agreement were transformed into percentages of the corresponding gait parameter mean values. In order to measure the test-retest repeatability, intra-class correlation coefficients ICC (3, 1) were calculated. A classification of ICC values according to [64] is as follows: 0.0 - 0.5: poor, 0.5 - 0.75: moderate, 0.75 - 0.90: good, 0.90 - 1.0: excellent.

## Acknowledgements

The authors would like to thank Keri Hartman for proofreading and linguistic advice.

## Conflicts of Interest

S. M. is an employee of Lindera, A. A. and S. R. were employees of Lindera, I. S., and A. S. declare no conflict of interest.